# INITIAL STUDY ON THE SHAPE OPTIMISATION OF THE CLIC CRAB CAVITY

P. K. Ambattu, G. Burt, R. G. Carter, A. C. Dexter, Cockcroft Institute, Lancaster University, Lancaster, UK, LA1 4YR

R.M. Jones, Cockcroft Institute, Manchester University, Warrington, UK, WA4 4AD

P. McIntosh, STFC, Daresbury Laboratory, Warrington, Cheshire, WA4 4AD, UK

*Abstract*

The compact linear collider (CLIC) requires a crab cavity to align bunches prior to collision. The bunch structure demands tight amplitude and phase tolerances of the RF fields inside the cavity, for the minimal luminosity loss. Beam loading effects require special attention as it is one potential sources of field errors in the cavity. In order to assist the amplitude and phase control, we propose a travelling wave (TW) structure with a high group velocity allowing rapid propagation of errors out of the system. Such a design makes the cavity structure significantly different from previous ones. This paper will look at the implications of this on other cavity parameters and the optimisation of the cavity geometry.

## INTRODUCTION

Deflection cavities were first proposed in 1988 for the rotation of particle bunches prior to the interaction point (IP) in the presence of a finite crossing angle. Efficient design of the crab cavity will lead to a flexible design of the interaction region. Unlike an accelerating cavity, a crab cavity is operated in the lowest dipole mode, which is $TM_{110}$-like in nature, as it has the highest transverse geometric shunt impedance defined as

$$\left(\frac{R}{Q}\right)_{\perp} = \frac{|V_z|^2}{\omega U}\left(\frac{c}{\omega r}\right)^2 \qquad (1)$$

where U is the stored energy and $V_z$ is the longitudinal voltage measured at the radial offset of r from the cavity axis at the angular frequency of ω [1]. For a crossing angle of $\theta_c$, the peak transverse kick voltage required is given by

$$V_{\perp} = \frac{\theta_c E_{beam} c}{2\omega R_{12}} \qquad (2)$$

where $E_{beam}$ is the beam energy and $R_{12}$ is the transfer matrix element representing the final focussing system, given by,

$$\Delta x_{ip} = R_{12} x_c' \qquad (3)$$

where $\Delta x_{ip}$ is the vertical displacement of the bunch at the IP and $x_c$' is the angular direction of the bunch when leaving the cavity [2]. Eq. (2) shows that higher frequency operation of the cavity reduces the kick required to produce a given deflection angle. For 1.5 TeV beams crossing at 20 mrad and for $R_{12} = 25$ m, the maximum kick voltage required is 2.4 MV. For a transverse gradient of 20 MV/m, a structure length of 120 mm would be required to achieve the above kick. For a phase advance of 2π/3 per cell, this requires about 15 cells. In addition to providing a kick at the operating frequency, the cavity should also suppress the wakefields that include monopole and higher order dipole modes in the cavity. However this aspect is not discussed here.

## CAVITY CELL SHAPE

The cavity cell shape is shown in Fig. 1. This is adopted from the basic disc-loaded waveguide structure, where the coupling between adjacent cells is achieved through a common iris. In the figure, L, R and C stands for the length, radius and curvature respectively and suffixes c, e and i stand for the cell, equator and iris respectively. The cell length (cavity + iris), is set by the phase advance per cell, $\Phi = k_{rf}(2L_c)$, where $k_{rf}$ is the free space phase constant in rad./meter. The iris edges are curved to avoid surface field enhancement and consequent field emission.

We studied the eigenmode characteristics of the single cell of an infinitely periodic copper structure as it can be used to infer the performance of a multi-cell structure. A $\Phi = 2\pi/3$ per cell structure, for which $2L_c = 8.337$ mm is modelled in Microwave Studio™ [3] as a first step.

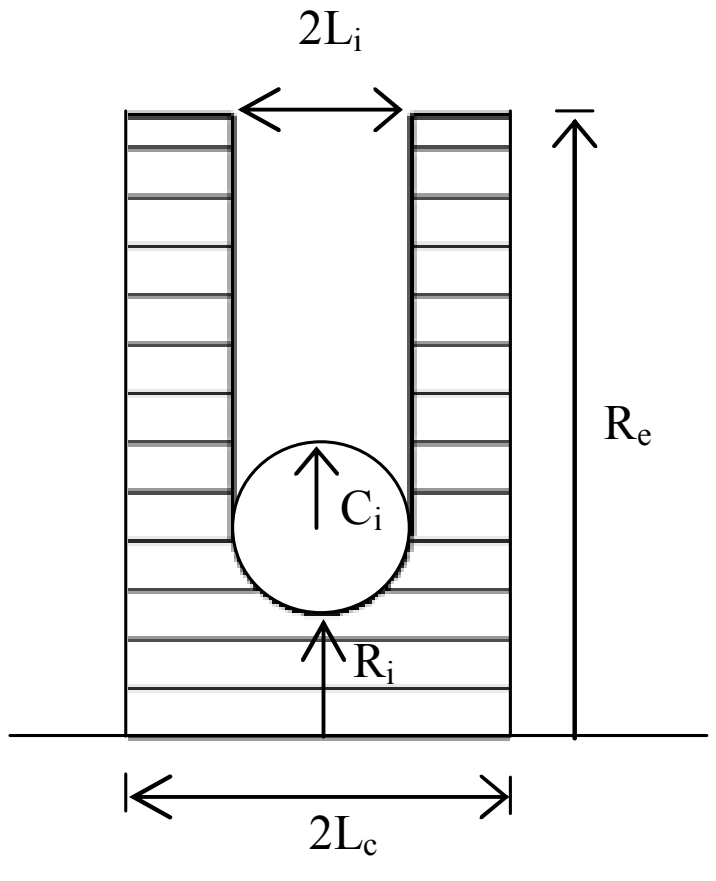

Figure 1: Cavity cell shape with dimensions

Various cell shapes are simulated for the fixed operating frequency and the important figures of merit such as the transverse R/Q, group velocity and maximum surface fields are calculated.

## RESULTS

The frequency sensitivity of the dipole cavity to its dimensions is shown in table 1, which demand high fabrication tolerances for the dipole cavity. Fig. 2 shows the transverse R/Q for various cell shapes. For lower iris radii and lengths, R/Q is almost a constant with the iris radius, but falls as the iris dimensions increase.

Table 1: Frequency sensitivity of the dipole cavity

| **Dimension (mm)** | **Freq. sensitivity (MHz/mm)** |
|---|---|
| $R_i$ | –231 |
| $R_e$ | –791 |
| $2L_i$ | –212 |
| $2L_c$ | 71 |

A maximum value around 4500 Ω/m is achieved at an iris radius of 2 mm and length of 1 mm. For the shunt impedance ≥ 67 % of the maximum value, the iris length is to be kept ≤ 75 % of the cell length. However, the minimum iris length is governed by good mechanical stability and effective heat removal.

Group velocities are calculated from the dispersion curves for a given cell shape and are shown in Fig. 3. A maximum value near 3 % of the velocity of light can be obtained with an iris radius of 4.25 mm and length of 1 mm. This is within the higher limit preferred for the X-band linacs [4]. This is because high group velocities enhance the impedance matching between the RF power and the breakdown spot. Also for a given iris radius, $v_{gr}$ varies inversely as the iris length. Also it was found that the cell with iris radii higher than 4.25 mm causes a coupling between the lower and upper dipole bands, hence changing the slope of the dispersion curve, thereby reducing the group velocity. The geometric dependence of group velocity is highly useful for the design of constant gradient TW structures [4] as long as its sign is not changed.

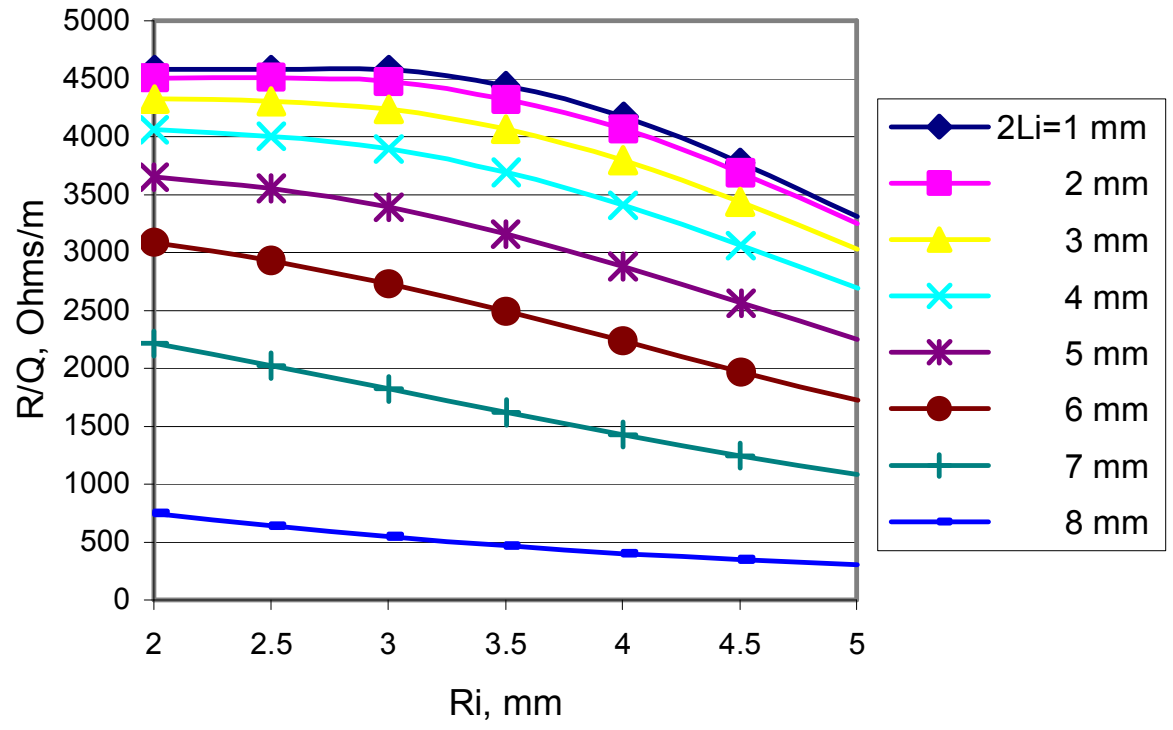

Figure 2: Iris radius × R/Q for different iris lengths

## RF FIELD ERRORS AND BEAM LOADING

Amplitude and phase errors are unavoidable in any RF system, which for a crab cavity result in less efficient collisions. The amplitude error causes a residual crossing angle at the IP while the phase error causes a transverse offset of the bunch center at the IP. For the crab crossing scheme, these tolerances are given by,

$$\frac{\Delta V}{V} < \frac{2\sigma_x^*}{\sigma_z \theta_c}\sqrt{\frac{1}{S^2}-1} \qquad (4)$$

$$\Delta\Phi < \frac{4\omega_{rf}\sqrt{\sigma_x^* \ln(1/S)}}{c\,\theta_c} \qquad (5)$$

where $\sigma_x^*$ is the horizontal IP beam size, $\sigma_z$ is the bunch length and S is the luminosity reduction factor [2]. For 2 % luminosity loss (S = 0.98) and a crossing angle of 20 mrad, the CLIC scheme (11.994 GHz, $\sigma_x^*$ = 40 nm, $\sigma_z$ = 45 µm) requires the above tolerances as 1.8 % and 0.016° for the amplitude and phase respectively. For the ILC crab cavity (3.9 GHz, $\sigma_x^*$ = 640 nm, $\sigma_z$ = 300 µm) the values are obtained as 4.33 % and 0.085° and for the NLC cavity (11.424 GHz, $\sigma_x^*$ = 243 nm, $\sigma_z$ = 110 µm) these would be 4.49 % and 0.095° respectively. Thus the CLIC crab cavity necessitates tighter amplitude and phase control system than the existing ones which recommends a TW structure with a high group velocity since it allows rapid extraction and correction of the RF errors.

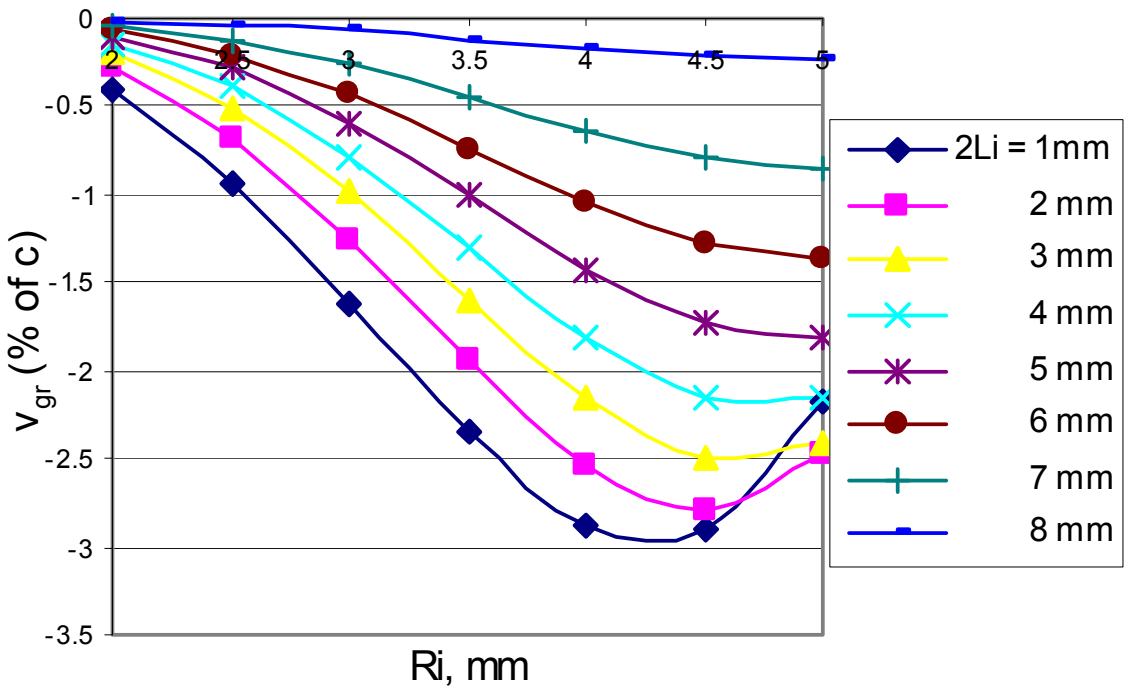

Figure 3: Iris radius × group velocity for different iris lengths

In a loaded cavity, the total field that the beam sees is the superposition of the generator induced and beam induced fields. For an ideal crab cavity, the fundamental component of the bunch current and the longitudinal cavity field ($E_z$) are in phase. If the beam doesn't arrive in phase with the cavity field, then a phase change will be introduced by the beam induced voltage. Also, the beam loading can cause amplitude errors; hence will be an assured source of the above stated errors in the CLIC crab cavity. The short bunch spacing (0.5 ns) of the CLIC structure makes the errors grow along the train (156 ns) causing displacement and/or uneven rotation of the bunches at IP. For a dipole cavity, the steady state beam loading is directly proportional to the ratio of geometric shunt impedance (R/Q) to the group velocity, the vertical offset from cavity center and the structure length. A lower $R/(Qv_{gr})$ can be realised with bigger coupling irises between the cells, to lower the beam loading in a constant impedance structure. In a constant gradient structure where the group velocity is lowered downstream along the structure, the beam loading may become even worse than that in a constant impedance structure.

## PEAK SURFACE FIELDS

In normal conducting cavities, the achievable peak surface electric field ($E_{max}$) is limited by field emission which can eventually lead to surface melting and breakdown. For such structures, the $E_{max}$ is expected to be around 100 MV/m which was the threshold of X-ray emission in normal conducting X-band TW cavity [5]. Peak surface magnetic field ($H_{max}$) is set by the limit at which the pulsed temperature rise on the structure (irises) is at around $40^{o}$ K. This is the threshold for forming micro cracks in OFE copper [6]. Since the peak fields are proportional to the gradient ($E_{trans}$), the quantities $E_{trans}/E_{max}$ and $H_{max}/E_{trans}$ are of importance and are plotted in Figs. 5 and 6, and can be considered as the normalised peak fields which depend only on the cell shape. As it is desired for the cavity to have lower surface fields at a given operating gradient, the $E_{trans}/E_{max}$ should be higher while the $H_{max}/E_{trans}$ should be lower. Though the surface fields are directly related to the iris dimensions, present results show that the fields have a higher dependence on the iris length than on the iris radius over the range of study. For the present dipole cavity, $E_{max}$ is concentrated on the equator faces (about half way on the $R_e$) while $H_{max}$ is concentrated on the iris. This makes the iris the hottest region in the cavity and should be provided with proper cooling channels.

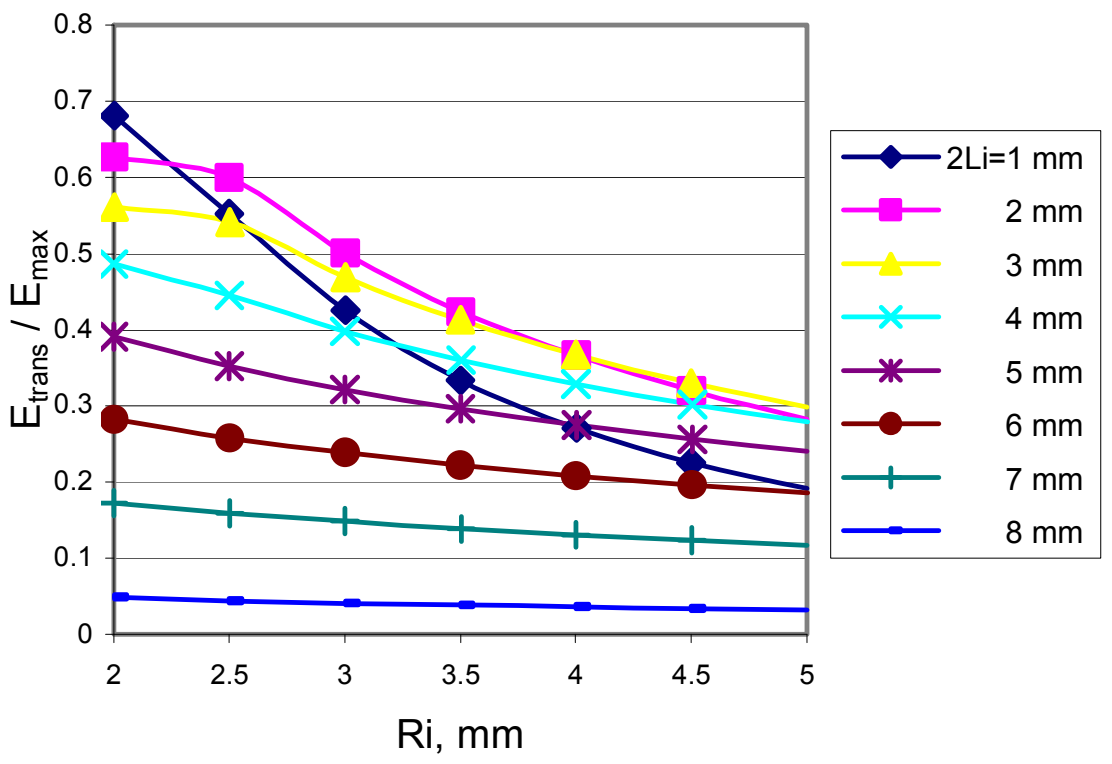

Figure 5: Iris radius × $E_{trans}/E_{max}$ for different iris lengths

Assuming a 20 MV/m gradient structure, the peak surface fields for the present $2\pi/3$ cell were evaluated. It was found that $E_{max}$ was less than 100 MV/m for iris radii less than 4.5 mm and lengths below 7 mm. Similarly, the peak surface heating was below $40^{o}$ K for all iris radii with iris lengths below 8 mm. Further simulations showed that the peak fields can be drastically reduced if longer (higher phase advance) cells are used, which allows the use of longer irises. However this will reduce the group velocity.

Considering the above discussion, a range of 4 to 5 mm for the iris radius and 2 to 4 mm for the iris length for the $2\pi/3$ phase advance/cell structure could be a preferred choice.

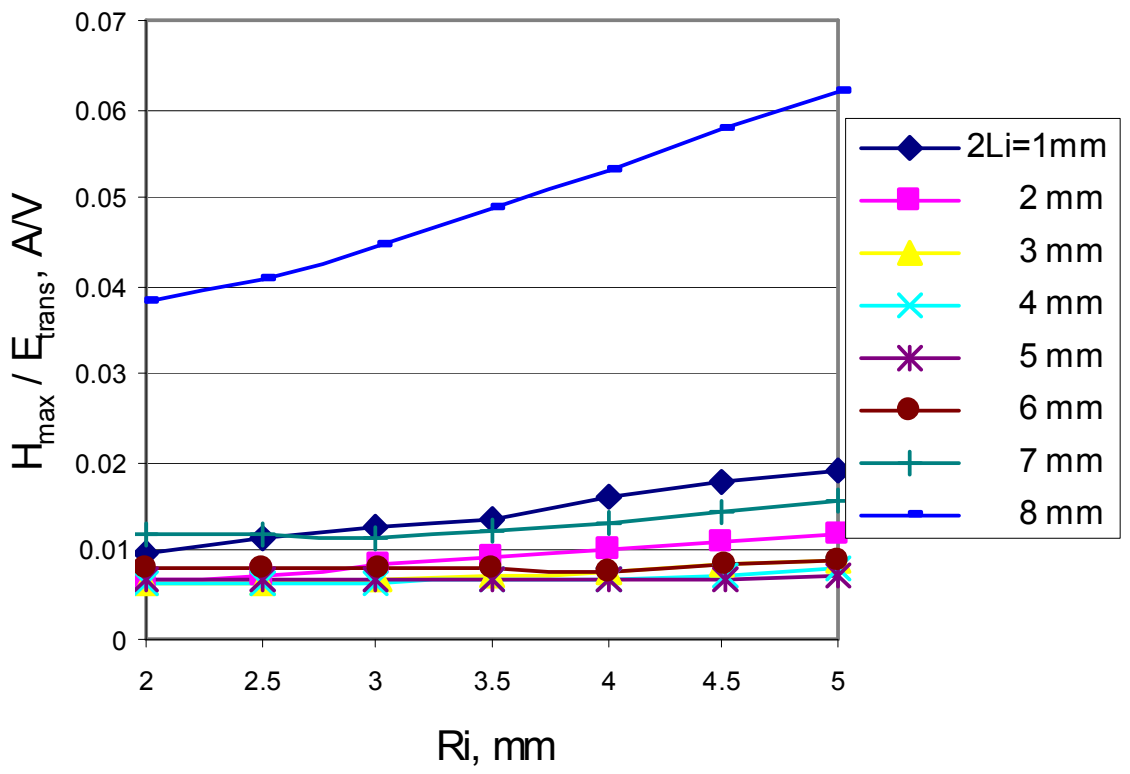

Figure 6: Iris radius × $H_{max}/E_{trans}$ for different iris lengths

## CONCLUSION

The behaviour of a normal conducting cavity is highly sensitive to the cell shape when operated in the X-band. For a given gradient and a desired net kick, the cell shape can be optimised by the proper selection of the cell profile. The key parameters of concern are the geometric shunt impedance, group velocity and surface fields. A TW structure with a low shunt impedance and high group velocity is suggestive for reduced beam loading and efficient RF control. This can be achieved by using bigger coupling irises as long as the surface fields are in the safe limits.

## ACKNOWLEDGEMENT

The work is supported by the Science and Technology Facilities Council (STFC).